\documentclass[letter,traditabstract]{aa}
\usepackage{subfigure}                                   
\usepackage{graphicx}
\usepackage{txfonts}
\begin{document}
%
   \title{Finding binary active galactic nuclei candidates by the centroid shift in imaging surveys\\
    II. Testing the method with SDSS J233635.75-010733.7}

   \author{Yuan Liu
          \inst{1}
          }

   \institute{Key Laboratory of Particle Astrophysics, Institute of High
Energy Physics, Chinese Academy of Sciences, P.O.Box 918-3, Beijing
100049, China\\
              \email{liuyuan@ihep.ac.cn}}

   \date{...}


  \abstract
   {In Liu (2015), we propose selecting binary active galactic nuclei (AGNs) candidates using the centroid shift of the images, which is induced by the non-synchronous variations of the two nuclei. In this paper, a known binary AGN (SDSS J233635.75-010733.7) is employed to verify the ability of this method. Using  162 exposures in the $R$ band of \textit{Palomar Transient Factory} (PTF), an excess of dispersion in the positional distribution of the binary AGN is detected, though the two nuclei cannot be resolved in the images of PTF. We also propose a new method to compare the position of the binary AGN in PTF $g$ and $R$ band and find the difference is highly significant even only with 20 exposures. This new method is efficient for two nuclei with different spectral energy distributions, e.g., type I + type II AGN or off-set AGN. Large-scale surveys, e.g., the Panoramic Survey Telescope and Rapid Response System and the Large Synoptic Survey Telescope, are expected to discover a large sample of binary AGN candidates with these methods.}

   {}

   \titlerunning{Finding binary AGNs by the centroid shift}
   \maketitle
%

\section{Introduction}
Supermassive black holes (SMBHs) are discovered in the majority of massive galaxies (Kormendy \& Richstone 1995). They should have grown through  mergers and gas accretion (Volonteri et al. 2003; Di Matteo et al. 2008; Kormendy \& Ho 2013). Thus, binary active galactic nuclei (AGNs) \footnote{In this paper, binary AGNs generally stand for any double sources discovered by imaging or spectroscopy. We do not strictly distinguish binary AGNs from dual AGNs or AGN pairs. } are expected to be common, since galaxy mergers can trigger the activity of SMBHs (Begelman et al. 1980; Hernquist 1989; Kauffmann \& Haehnelt 2000; Hopkins et al. 2008). Although hundreds of binary AGNs are found with separations of tens of kiloparsec (kpc), only about ten binary AGNs are known with separations $\sim$10 kpc (Junkkarinen et al. 2001; Komossa et al. 2003; Hennawi et al. 2006; Fu et al. 2011; Koss et al. 2011; Mazzarella et al. 2012; Liu et al. 2013; M{\"u}ller-S{\'a}nchez et al. 2015). The occurrence rate of kpc-scale binary AGNs ($\sim$1\%) is lower than the expected value by a factor of ten if each major merger can induce a binary AGN (Yu et al. 2011). Non-simultaneous activity and the gas content of galaxies are invoked to reconcile this apparent discrepancy (Foreman et al. 2009; Yu et al. 2011; Van Wassenhove et al. 2012).

On the observational aspect, a large and complete sample is important for the assessment of  this discrepancy. Serendipitously discovered binary AGNs are rare and not sufficient to build a statistically meaningful sample. A systematic method is required for identifying kpc-scale binary AGN candidates.  Double-peaked narrow lines (e.g., double-peaked [O III] lines) have been utilized to select kpc-scale binary AGNs (Wang et al. 2009; Liu et al. 2010; Smith et al. 2010; Shen et al. 2011). However,  double-peaked narrow lines are usually produced by the gas dynamics (Fu et al. 2012; Blecha et al. 2013; M{\"u}ller-S{\'a}nchez et al. 2015).
 Moreover, the line shift of a binary AGN in a face-on orbit is small and undetectable.
Therefore, an efficient method is needed to well complement spectroscopy methods.

In Liu (2015, hereafter Paper I), we show that the imaging centroid of a binary AGN will shift due to the non-synchronous variation of the two nuclei. Thus, such binary AGNs can be revealed by multi-epoch observations, even if the separation is smaller than the angular resolution. This method utilizes  the violent variation of AGNs; thus, it is more suitable for two type I AGNs or blazars with strong variability and its efficiency is low for type I + type II binaries. Therefore, we still need an efficient method for the pairs including type II AGN or a galactic nucleus without an AGN. The continuum of type I AGN is bluer than type II AGN or a galactic nucleus without an AGN, which is another distinguishable property of AGNs that can be used to uncover binary AGNs. If the spectral energy distributions (SEDs) of two nuclei are significantly different and two or more filters are available, the images of the blue and red filter are more dominated by the bluer and redder nucleus respectively. As a result, the centroid of the binary in the blue (red) filter should shift towards the bluer (redder) nucleus compared with the centroid in the red (blue) filter. This new method should be valid for two nuclei with different SEDs, e.g., type I + type II AGN, type I AGN+ broad absorption line (BAL) quasar, and type I AGN + galaxy (offset AGN). It is also appropriate to two type II nuclei if the colors of the two host galaxies are remarkably different. As we mentioned in Paper I, the fast variation in jet or the reflection from a cloud near a single AGN can mimic another ``nuclei'' in the image and thus contaminates the candidates selected by the centroid shift. Follow-up high resolution imaging and spectroscopy are still required to confirm the candidates. Therefore, before hunting for new binary AGNs, we would like to select a known binary AGN to verify the two methods.

 In Sect. 2, we describe the target and data selection. In Sect. 3, the procedure for calculation of the centroid in one band is explained. In Sect. 4, the result of the centroid shift between two bands is shown. In Sect. 5, we discuss the implications of the results and present our conclusions.


\section{Target and Data Selection}

 \begin{figure*}[!t]
 \centering
   \includegraphics[width=0.90\textwidth]{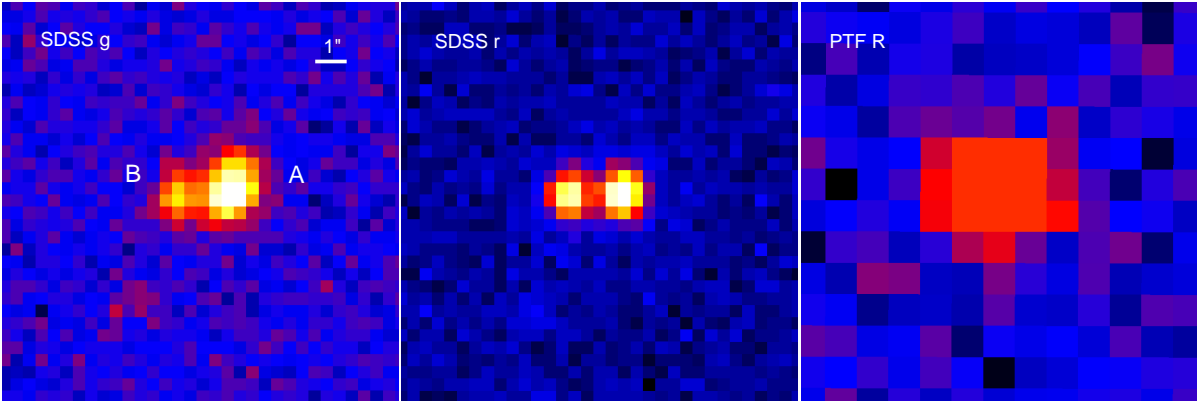}
   \caption{Images of SDSS J233635.75-010733.7 in SDSS $g$ band (\textit{left}),SDSS $r$ band (\textit{middle}), and PTF $R$ band (\textit{right}). The two nuclei are well resolved in the SDSS images (quasar A is a normal quasar and quasar B is a BAL quasar), but they are blended in the PTF image due to the relatively poor seeing and large pixel scale.}
 \label{img}
 \end{figure*}

Since more than 100 exposures are required by the method proposed in paper I,  the \textit{Palomar Transient Factory} (PTF), with a large sky coverage and frequent exposures, is selected to test our methods. There are three criteria employed to select a known binary AGN covered by PTF: (1) there is at least one type I nucleus in the binary; (2) the two nuclei are not resolved in PTF images; (3) there are more than 100 exposures to ensure the significance. As a result,  only one source, SDSS J233635.75-010733.7, is found to be suitable for testing our method. This source is discovered by an infrared imaging survey at the Keck Observatory and confirmed to be a binary quasar (a standard quasar with a blue continuum and broad emission lines and a BAL quasar) by resolved optical spectra (Gregg et al. 2002). The separation of the two quasars is $1.67''$, while the mean seeing of PTF observations is $2.1''$ and $2.2''$ in $R$ and $g$ band, respectively. The pixel scale of PTF CCD is $1.01''$ (Law et al. 2009). Thus, the two quasars cannot be resolved in the image of PTF (Figure \ref{img} \textit{right}), though the source slightly deviates from the point spread function (PSF). PTF observed this source 162 and 103 times at $R$ and $g$ band, respectively\footnote{http://irsa.ipac.caltech.edu/applications/ptf/}. As a result, this source is appropriate for the test on the centroid method.

\section{Centroid Shift in $R$ band}
We present a realistic procedure for the method proposed in Paper I in this section for the $R$ band exposures of SDSS J233635.75-010733.7.

The centroids in pixel units of all sources in the 162 $R$ band exposures were calculated by \texttt{SExtractor}\footnote{http://www.astromatic.net/software/sextractor}. The sources with a significance higher than $3\sigma$ in at least 4 adjacent pixels were extracted. The windowed centroids, XWIN\_ IMAGE and YWIN\_IMAGE, were  adopted to achieve the accuracy close to the theoretical limit set by image noise\footnote{https://www.astromatic.net/pubsvn/software/sextractor/trunk\\/doc/sextractor.pdf}. The RA and Dec in J2000 were also recorded.

A master frame was selected to define the baseline of positions. Since we only need relative astrometry, the selection of the master frame is arbitrary and our final result is not sensitive to it. In practice, the exposure with a high signal-to-noise ratio (SNR) is preferred to include more reference sources.

The sources around SDSS J233635.75-010733.7 with  distances between $20''$ and $180''$ were identified as  reference sources. In total, there are 18 reference sources. However, some of them may be not available in some exposures with low SNRs. The centroid of the target quasar in the master frame is noted as $X^q$ and $Y^q$, and the centroids of the reference sources in the master frame are $X^r_i$ and $Y^r_i$, where $i=1-18$.

\begin{figure*}[]
 \centering
   \includegraphics[width=0.48\textwidth]{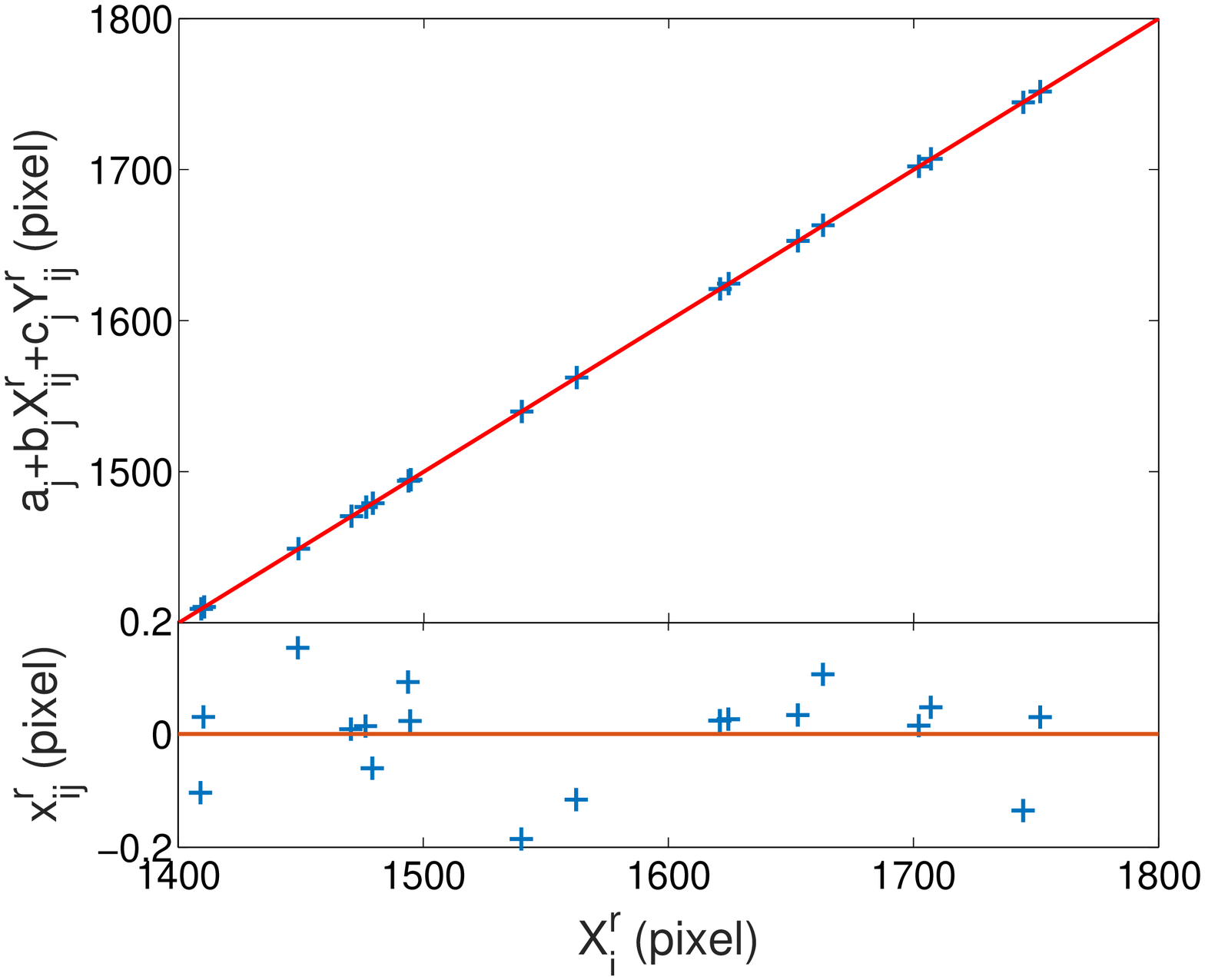}
   \includegraphics[width=0.495\textwidth]{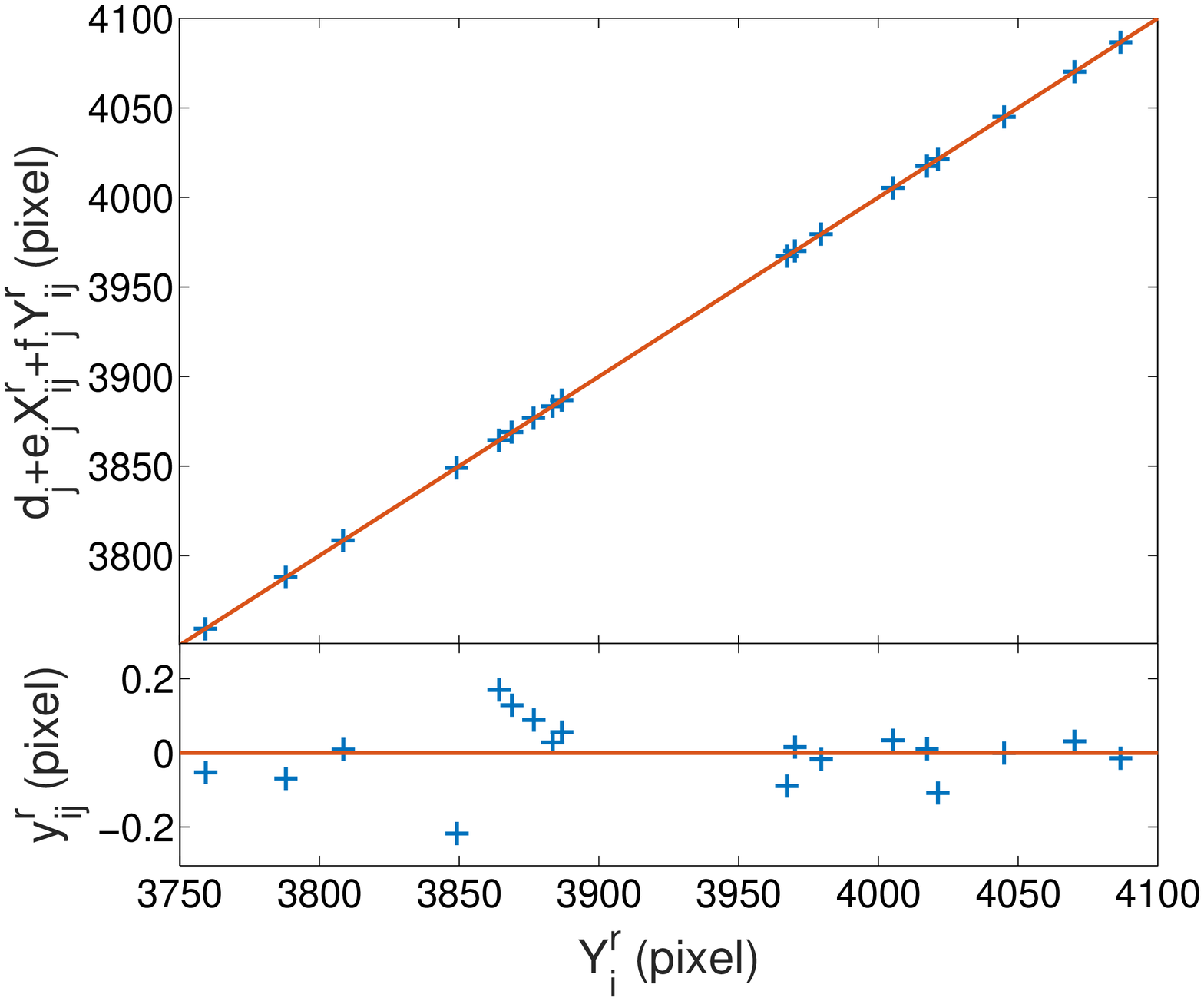}
   \caption{An example of the linear fitting to determine the transformation coefficient between the master frame and a given frame. The residuals of $x$ and $y$  are well scattered around zero under the linear transformation; therefore, no higher order term is included in the current work.}
 \label{xy_res}
 \end{figure*}

The centroids of the detected reference sources are adopted to determine the coordinate transformation between every frame and the master frame. A linear transformation is sufficient for  small angular scales, i.e.,

\begin{eqnarray}
  X^r_i & = & a_j+b_jX^r_{ij}+c_jY^r_{ij} \ ,\\
  Y^r_i & = & d_j+e_jX^r_{ij}+f_jY^r_{ij}  \ ,
\end{eqnarray}
where $X^r_{ij}$ and $Y^r_{ij}$ are the centroids of the \textit{i}th reference source in \textit{j}th frame (excluding the master frame).
If the detected reference sources are more than six in one frame, we can fit Eq. (1) and (2) to determine the transformation coefficients $a_j\cdots f_j$ of this frame. The frames without sufficient reference sources were discarded.  Figure \ref{xy_res} shows an example of the linear fitting.

The residual of the position of reference sources after the transformation determines the astrometric accuracy, i.e.,
\begin{eqnarray}
  x^r_{ij}=X^r_i-(a_j+b_jX^r_{ij}+c_jY^r_{ij}) \ ,\\
  y^r_{ij}=Y^r_i-(d_j+e_jX^r_{ij}+f_jY^r_{ij}) \ .
\end{eqnarray}

We then calculated the residual of the position of the target quasar in \textit{j}th frame using the same transformation, i.e.,
\begin{eqnarray}
  x^q_{j}=X^q-(a_j+b_jX^q_{j}+c_jY^q_{j}) \ ,\\
  y^q_{j}=Y^q-(d_j+e_jX^q_{j}+f_jY^q_{j}) \ ,
\end{eqnarray}
where $X^q_{j}$ and $Y^q_{j}$ are the  centroids of the target quasar in \textit{j}th frame (exclude the master frame).

We removed the outliers in $x^r_{ij}$, $y^r_{ij}$,  $x^q_{j}$, and $y^q_{j}$ that deviated from the mean value larger than $4 \sigma$. This process was performed iteratively until no new outlier is found.
Then the systematic offset in $x^q_{j}$ and $y^q_{j}$ was corrected, i.e., the means of $x^q_{j}$ and $y^q_{j}$ were set to be zero.

The distribution of $x^r_{ij}$, $y^r_{ij}$,  $x^q_{j}$, and $y^q_{j}$ should be similar if the target quasar is a single AGN. However, the distribution of the residual of a binary AGN should be elongated along the direction of the two nuclei (see the simulations in Paper I). A large astrometric error or a small sample could hide the elongated distribution.
Since the direction of the two nuclei is not known a priori, we performed the principle component analysis on $x^q_{j}$ and $y^q_{j}$ to find the direction of the maximum dispersion.

Then $x^r_{ij}$, $y^r_{ij}$,  $x^q_{j}$, and $y^q_{j}$ were projected to the direction of the maximum dispersion and the perpendicular direction. The corresponding distributions are noted as $\tilde{x}^r$, $\tilde{y}^r$,  $\tilde{x}^q$, and $\tilde{y}^q$.

The  Kolmogorov-Smirnov (K-S) test is utilized to examine whether there is significant difference between these distributions. The  distribution $\tilde{x}^q$ is expected to be different from the others.
We performed K-S tests on three pairs, i.e., $\tilde{x}^q$ vs. $\tilde{x}^r$, $\tilde{x}^r$ vs. $\tilde{y}^r$,  and $\tilde{y}^q$ vs. $\tilde{y}^r$ and the p-values are 0.0075, 0.77, and 0.076, respectively. Thus the distribution of the residual of the quasar position is indeed different from that of the reference sources in one direction. Furthermore, the standard deviations of $\tilde{x}^q$,  $\tilde{y}^q$, $\tilde{x}^r$, and $\tilde{y}^r$ are 0.125, 0.099,  0.085, and 0.089 pixels,respectively, which are consistent with the result of K-S tests.

 \begin{figure}[!htp]
 \centering
   \includegraphics[width=0.45\textwidth]{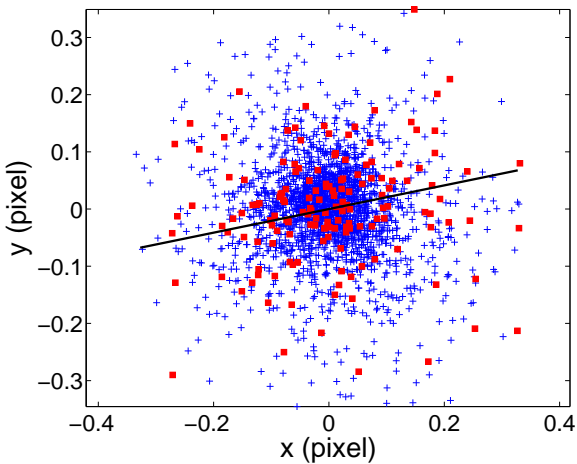}
   \caption{Distributions of the residual of the position in PTF $R$ band. Squares are the residual of the quasar ($x^q$  and $y^q$); pluses are the residual of the reference sources ($x^r$ and $y^r$). The solid line indicates the direction of $\tilde{x}^q$.}
 \label{r_res}
 \end{figure}

 \begin{figure}[!htp]
 \centering
   \includegraphics[width=0.45\textwidth]{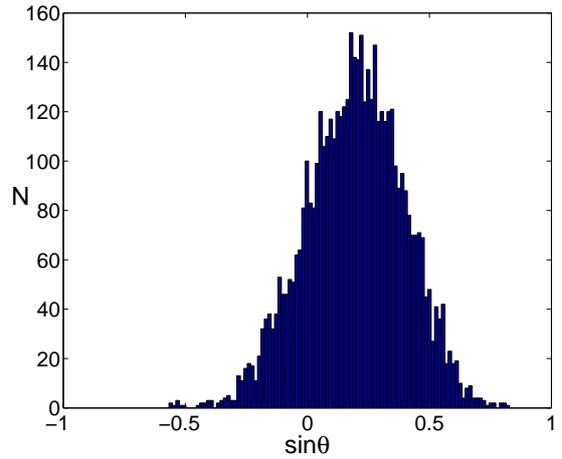}
   \caption{Distribution of the direction of $\tilde{x}^q$ from 5000 bootstrap runs (see the text in Sect. 3 for details).}
 \label{theta}
 \end{figure}

Figure \ref{r_res} shows the distribution of $\tilde{x}^r$, $\tilde{y}^r$,  $\tilde{x}^q$, and $\tilde{y}^q$ and the direction of $\tilde{x}^q$. We performed 5000 bootstrap runs from $x^q$ and $y^q$ to estimate the error of the direction of $\tilde{x}^q$. The angle between the $x$-axis  and $\tilde{x}^q$ is noted as $\theta$. Figure \ref{theta} shows the distribution of $\sin\theta$. The mean of $\sin\theta$ is 0.20 and the standard deviation is 0.20. The actual direction of the two nuclei is $\sin\theta=0.03$; thus, the direction of $\tilde{x}^q$ is consistent with it at $1\sigma$ level.

\section{Centroid Shift between $R$ and $g$ band}
The method presented in Sect. 3 is appropriate for AGNs with strong variability, i.e., type I AGNs or blazars. In this section, we will employ the exposures of $R$ and $g$ band at the same time to  alleviate this requirement. According to the optical spectra and  SEDs of the two quasars in  SDSS J233635.75-010733.7 (Figure 2 in Gregg et al. 2002), the fluxes of the two quasars are comparable in $R$ band, but the flux of the BAL quasar in $g$ band is lower than that of the blue quasar by a factor of 5 due to its intrinsic absorption. Thus, the distribution of the centroids in the two bands should be different. There are only $R$ and $g$ filters in PTF, though the filters with a larger spectral separation could promote the  significance.

 \begin{figure}[!htp]
 \centering
   \includegraphics[width=0.45\textwidth]{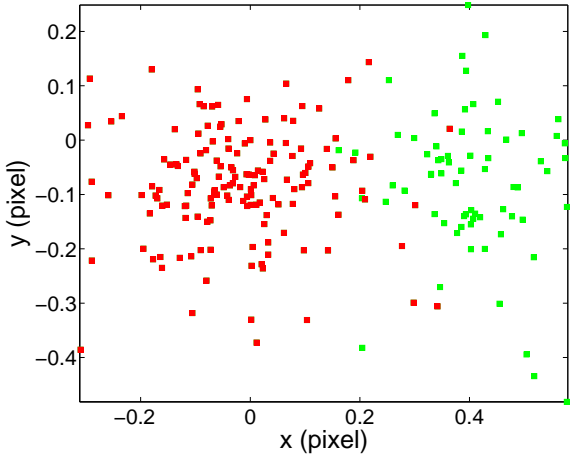}
   \caption{Comparison between the distribution of the residual of the position in PTF $R$ (red squares) and $g$ (green squares) band. The offset relative to the origin (defined by the mean of the residual of the reference sources) is not corrected, since it will not influence the relative positions of the residuals of the two bands. }
 \label{r_g_res}
 \end{figure}

The same procedures in Sect. 3 were performed on $g$ band using the same reference sources identified in the master frame of the $R$ band.

Figure \ref{r_g_res} compares the distributions of the residuals in $R$ and $g$ band. Since the direction of the two nuclei is nearly parallel to the $x$ direction, the residuals in $g$ and $R$ band are expected to be only different in the $x$ direction. The significance indicated by K-S test between the residual in $g$ and $R$ band is $1.6\times10^{-39}$ in the $x$ direction; while the significance is only 0.42 for the $y$ direction. This result is consistent with the visual inspection.

\section{Discussion and Conclusions}
Using a confirmed binary AGN, we have tested the validity of the two centroid methods. Both of the two methods have revealed the signal of two nuclei.  The significance of the method using one filter is relatively low (p-value=0.0075), since the variation of the BAL nucleus is not as strong as the type I nucleus. For the method using two filters, the distributions of the residual in $R$ and $g$ band are significantly different (p-value=$1.6\times10^{-39}$)  and consistent with the expectation from the SEDs of the two quasars. Actually, if we randomly select 20 exposures of $R$ and $g$ band respectively, the significance of the difference between the distributions of $R$ and $g$ band is still at $10^{-9}-10^{-6}$ level. These results indicate the success of the centroid method in uncovering the known binary AGN and its potential power in discovering new binary AGNs.

During the merging process of galaxies, the enhanced inflow may induce strong star formation and significant obscuration. Thus type I+type II or type I+BAL cases should be more common than the type I+type I case (Sanders et al. 1988; Treister et al. 2010). The so-called ``offset" AGNs selected by the multi-filter method are also thought to be the important tracer of the merging process and valuable for the understanding on the fueling of AGNs (Comerford \& Greene2014; Steinborn et al. 2015).  However, the multi-filter method is not helpful for two type I or type II AGNs with similar SEDs.

For the multi-filter method, the relative astrometric error between two bands is crucial. Pier et al. (2003) calibrated the astrometry of SDSS images  and found that the standard deviation of the distribution of position differences between $r$ filter and other filters is about $0.025''$, which is better than the astrometric error of $r$ filter against the USNO CCD Astrograph Catalog by a factor of 2. Our method only needs the relative astrometry in arcmin scale and thus the accuracy should be much better. Actually, the distributions of $\tilde{x}^r$ and $\tilde{y}^r$ of $R$ and $g$ band are well consistent with each other.

{A binary system such as SDSS J233635.75-010733.7 is hard to discover by optical spectroscopy due to its high redshift ($z=1.285$). The normally employed lines, e.g.,  [O III] and H$\beta$ lines, are already shifted into the infrared band. A large scale and deep infrared spectroscopy survey  on AGNs is still not available at present. However, the centroid methods proposed here are not limited by the redshift in this content if the astrometric error is small enough to identify the centroid shift.

The multi-filter method does not require strong variations. Thus, even the exposures in the same night are helpful for identifying the relative shift between different bands. A large sample of binary AGNs should be quickly established by large-scale surveys, e.g., the Panoramic Survey Telescope and Rapid Response System and the Large Synoptic Survey Telescope.


\begin{acknowledgements}
The author thanks the referee for useful comments that clarified the paper. Some data in this work were obtained as part of the \textit{Palomar
Transient Factory} (PTF) project.
This work is supported by the National Natural Science Foundation of
China under grant Nos. 11573027, 11103019, 11133002, and 11103022.
\end{acknowledgements}


\begin{thebibliography}{}

\bibitem[Begelman et al.(1980)]{1980Natur.287..307B} Begelman, M.~C.,
Blandford, R.~D., \& Rees, M.~J.\ 1980, \nat, 287, 307


\bibitem[Comerford
\& Greene(2014)]{2014ApJ...789..112C} Comerford, J.~M., \& Greene, J.~E.\ 2014, \apj, 789, 112




\bibitem[Di Matteo et al.(2008)]{2008ApJ...676...33D} Di Matteo, T.,
Colberg, J., Springel, V., Hernquist, L.,
\& Sijacki, D.\ 2008, \apj, 676, 33





\bibitem[Foreman et al.(2009)]{2009ApJ...693.1554F} Foreman, G., Volonteri,
M., \& Dotti, M.\ 2009, \apj, 693, 1554


\bibitem[Fu et al.(2011)]{2011ApJ...740L..44F} Fu, H., Zhang, Z.-Y., Assef,
R.~J., et al.\ 2011, \apjl, 740, L44


\bibitem[Fu et al.(2012)]{2012ApJ...745...67F} Fu, H., Yan, L., Myers,
A.~D., et al.\ 2012, \apj, 745, 67




\bibitem[Gregg et al.(2002)]{2002ApJ...573L..85G} Gregg, M.~D., Becker,
R.~H., White, R.~L., et al.\ 2002, \apjl, 573, L85



\bibitem[Hennawi et al.(2006)]{2006AJ....131....1H} Hennawi, J.~F.,
Strauss, M.~A., Oguri, M., et al.\ 2006, \aj, 131, 1


\bibitem[Hernquist(1989)]{1989Natur.340..687H} Hernquist, L.\ 1989, \nat,
340, 687



\bibitem[Hopkins et al.(2008)]{2008ApJS..175..356H} Hopkins, P.~F.,
Hernquist, L., Cox, T.~J., \& Kere{\v s}, D.\ 2008, \apjs, 175, 356



\bibitem[Junkkarinen et al.(2001)]{2001ApJ...549L.155J} Junkkarinen, V.,
Shields, G.~A., Beaver, E.~A., et al.\ 2001, \apjl, 549, L155


\bibitem[Kauffmann
\& Haehnelt(2000)]{2000MNRAS.311..576K} Kauffmann, G., \& Haehnelt, M.\ 2000, \mnras, 311, 576

\bibitem[Komossa et al.(2003)]{2003ApJ...582L..15K} Komossa, S., Burwitz,
V., Hasinger, G., et al.\ 2003, \apjl, 582, L15

\bibitem[]{} Kormendy, J., \& Richstone, D. 1995, ARA\&A, 33, 581


\bibitem[Kormendy
\& Ho(2013)]{2013ARA&A..51..511K} Kormendy, J., \& Ho, L.~C.\ 2013, \araa, 51, 511




\bibitem[Koss et al.(2011)]{2011ApJ...735L..42K} Koss, M., Mushotzky, R.,
Treister, E., et al.\ 2011, \apjl, 735, L42

\bibitem[Law et al.(2009)]{2009PASP..121.1395L} Law, N.~M., Kulkarni,
S.~R., Dekany, R.~G., et al.\ 2009, \pasp, 121, 1395

\bibitem[Liu et al.(2010)]{2010ApJ...715L..30L} Liu, X., Greene, J.~E.,
Shen, Y., \& Strauss, M.~A.\ 2010, \apjl, 715, L30

\bibitem[Liu et al.(2013)]{2013ApJ...762..110L} Liu, X., Civano, F., Shen, Y., et al.\ 2013, \apj, 762, 110

\bibitem[Liu(2015)]{2015A&A...580A.133L} Liu, Y.\ 2015, \aap, 580, A133


\bibitem[Mazzarella et al.(2012)]{2012AJ....144..125M} Mazzarella, J.~M., Iwasawa, K., Vavilkin, T., et al.\ 2012, \aj, 144, 125


\bibitem[M{\"u}ller-S{\'a}nchez et al.(2015)]{2015ApJ...813..103M}
M{\"u}ller-S{\'a}nchez, F., Comerford, J.~M., Nevin, R., et al.\ 2015,
\apj, 813, 103


\bibitem[Pier et al.(2003)]{2003AJ....125.1559P} Pier, J.~R., Munn, J.~A.,
Hindsley, R.~B., et al.\ 2003, \aj, 125, 1559

\bibitem[Sanders et al.(1988)]{1988ApJ...325...74S} Sanders, D.~B., Soifer,
B.~T., Elias, J.~H., et al.\ 1988, \apj, 325, 74



\bibitem[Shen et al.(2011)]{2011ApJ...735...48S} Shen, Y., Liu, X., Greene,
J.~E., \& Strauss, M.~A.\ 2011, \apj, 735, 48



\bibitem[Smith et al.(2010)]{2010ApJ...716..866S} Smith, K.~L., Shields,
G.~A., Bonning, E.~W., et al.\ 2010, \apj, 716, 866


\bibitem[Steinborn et al.(2015)]{2015arXiv151008465S} Steinborn, L.~K.,
Dolag, K., Comerford, J.~M., et al.\ 2015, arXiv:1510.08465


\bibitem[Treister et al.(2010)]{2010Sci...328..600T} Treister, E.,
Natarajan, P., Sanders, D.~B., et al.\ 2010, Science, 328, 600

\bibitem[Van Wassenhove et al.(2012)]{2012ApJ...748L...7V} Van Wassenhove,
S., Volonteri, M., Mayer, L., et al.\ 2012, \apjl, 748, L7


\bibitem[Volonteri et al.(2003)]{2003ApJ...582..559V} Volonteri, M.,
Haardt, F., \& Madau, P.\ 2003, \apj, 582, 559



\bibitem[Wang et al.(2009)]{2009ApJ...705L..76W} Wang, J.-M., Chen, Y.-M.,
Hu, C., et al.\ 2009, \apjl, 705, L76

\bibitem[Yu et al.(2011)]{2011ApJ...738...92Y} Yu, Q., Lu, Y., Mohayaee,
R., \& Colin, J.\ 2011, \apj, 738, 92


\end{thebibliography}
\end{document}